\documentclass[prd, twocolumn, showkeys, superscriptaddress, 9pt]{revtex4}
\usepackage{graphicx}
\usepackage{dcolumn}  
\usepackage{bm}          
\usepackage{xcolor}
\usepackage{amsfonts}
\usepackage{amsmath}
\newcommand{\beq}{\begin{equation}}
\newcommand{\eeq}{\end{equation}}

\begin{document}

\title{Ternary Quarter Wavelength  Coatings\\ for Gravitational Wave Detector Mirrors:\\Design Optimization via Exhaustive Search }
%
\author{V. Pierro}
\email[]{pierro@unisannio.it}
\affiliation{Dip. di Ingegneria, DING, Universit\'{a} del Sannio, I-82100 Benevento, Italy.}
\affiliation{INFN, Sezione di Napoli Gruppo Collegato di Salerno,  I-80126 Napoli, Italy.}
\author{V. Fiumara}
\affiliation{Scuola di Ingegneria, Universit\'{a} della Basilicata, I-85100 Potenza, Italy.}
\affiliation{INFN, Sezione di Napoli Gruppo Collegato di Salerno,  I-80126 Napoli, Italy.}
\author{F. Chiadini}
\affiliation{Dip. di Ingegneria Industriale, DIIN, Universit\'{a} di Salerno, I-84084 Fisciano, Salerno, Italy.}
\affiliation{INFN, Sezione di Napoli Gruppo Collegato di Salerno,  I-80126 Napoli, Italy.}
\author{V. Granata}
\affiliation{CNR-SPIN, c/o Universit\'{a} di Salerno, I-84084 Fisciano, Salerno, Italy.}
\affiliation{INFN, Sezione di Napoli Gruppo Collegato di Salerno,  I-80126 Napoli, Italy.}
\author{O. Durante}
\affiliation{Dip. di Fisica "E.R. Caianiello", Universit\'{a} di Salerno, I-84084 Fisciano, Salerno, Italy.}
\affiliation{INFN, Sezione di Napoli Gruppo Collegato di Salerno,  I-80126 Napoli, Italy.}
\author{J. Neilson}
\affiliation{Dip. di Ingegneria, DING,  Universit\'{a} del Sannio, I-82100 Benevento, Italy.}
\affiliation{INFN, Sezione di Napoli Gruppo Collegato di Salerno,  I-80126 Napoli, Italy.}
\author{C. Di Giorgio}
\affiliation{Dip. di Fisica "E.R. Caianiello", Universit\'{a} di Salerno, I-84084 Fisciano, Salerno, Italy.}
\affiliation{INFN, Sezione di Napoli Gruppo Collegato di Salerno,  I-80126 Napoli, Italy.}
\author{R. Fittipaldi}
\affiliation{CNR-SPIN, c/o Universit\'{a} di Salerno, I-84084 Fisciano, Salerno, Italy.}
\affiliation{INFN, Sezione di Napoli Gruppo Collegato di Salerno,  I-80126 Napoli, Italy.}
\author{G. Carapella}
\affiliation{Dip. di Fisica "E.R. Caianiello", Universit\'{a} di Salerno, I-84084 Fisciano, Salerno, Italy.}
\affiliation{INFN, Sezione di Napoli Gruppo Collegato di Salerno,  I-80126 Napoli, Italy.}
\author{F. Bobba}
\affiliation{Dip. di Fisica "E.R. Caianiello", Universit\'{a} di Salerno, I-84084 Fisciano, Salerno, Italy.}
\affiliation{INFN, Sezione di Napoli Gruppo Collegato di Salerno,  I-80126 Napoli, Italy.}
\author{M. Principe}
\affiliation{Dip. di Ingegneria, DING, Universit\'{a} del Sannio, I-82100 Benevento, Italy.}
\affiliation{INFN, Sezione di Napoli Gruppo Collegato di Salerno,  I-80126 Napoli, Italy.}
\affiliation{Museo Storico della Fisica e Centro Studi e Ricerche "Enrico Fermi", I-00184  Roma, Italy.}
\author{I. M. Pinto}
\affiliation{Dip. di Ingegneria, DING, Universit\'{a} del Sannio, I-82100 Benevento, Italy.}
\affiliation{INFN, Sezione di Napoli Gruppo Collegato di Salerno,  I-80126 Napoli, Italy.}
\affiliation{Museo Storico della Fisica e Centro Studi e Ricerche "Enrico Fermi", I-00184  Roma, Italy.}
%
\begin{abstract}
%
Multimaterial optical coatings are a  promising viable  option to meet the challenging requirements  
(in terms of transmittance, absorbance, and thermal noise) of next generation gravitational wave detector mirrors.
In this paper we focus on ternary coatings consisting of quarter-wavelength thick layers, 
where a third material ($H'$)  is added to the two presently in use, namely, silica ($L$) and titania-doped tantala ($H$),
featuring higher dielectric contrast (against silica) and lower thermal noise (compared with titania-doped tantala), 
but higher optical losses.
We seek the optimal {\it material sequences},  featuring minimal thermal (Brownian) noise under prescribed 
transmittance and absorbance constraints, by exhaustive simulation over {\it all} possible configurations, 
for different values of the optical density and extinction coefficient of the third material, including
the case of amorphous silicon and silicon nitride operating at ambient and cryogenic temperatures.
In all cases studied, the optimal designs consist of a stack of $(H'|L)$ doublets topped by a stack of $(H|L)$ doublets, 
confirming previous heuristic assumptions, and the achievable coating noise power spectral density reduction factor 
ranges from $\sim 0.5$ at $290\mbox{K}$  down to $\sim 0.1$ at $20\mbox{K}$. 
The robustness of the found optimal designs against layer thickness deposition errors and uncertainties and/or fluctuations
in the optical losses of the third material is also investigated. Possible margins for further thermal noise reduction by layer
thickness optimization, and strategies to implement it, are discussed.
\end{abstract}
%
%
\vspace{2pc}
\maketitle


\section{INTRODUCTION}
\label{sec:Intro}

The visibility distance of the currently operating interferometric detectors of gravitational waves is set
by thermal (Brownian) noise in the highly-reflective dielectric-multilayer coated  mirrors of their optical
cavities \cite{Martynov}.

During the last two decades, the quest for coating materials featuring large
dielectric contrast, low optical absorption (and scattering), and low mechanical losses 
(directly related to thermal noise, in view of the fluctuation-dissipation principle) 
has been quite intense.

Since the development in 1997 of titania-doped tantala 
($\mbox{TiO}_2 ::\mbox{Ta}_2\mbox{O}_5$) by Laboratoire des
Mat\'eriaux Avanc\'es  (LMA) \cite{Harry2007}, still used, together with
silica ($\mbox{SiO}_2$) , in the mirrors of the (currently operational)  ``advanced''  LIGO and Virgo detectors \cite{aLIGO,adVirgo}, several
potentially interesting coating materials have been investigated (see Ref. \cite{Vajente} for a recent review). 

None of  them qualifies as a  straight substitute for the materials currently in use, 
but a few of them are {\it better} in terms  of {\it some} properties (e.g., optical density, and/or mechanical losses), 
while unfortunately {\it worse} regarding others.

Amorphous silicon ({\it a}Si), in particular, has received much attention, in view of its large refractive index and
limited mechanical losses, down to cryogenic temperatures \cite{aSi}.
Its optical losses, on the other hand, are appreciably larger than those of titania-doped tantala, and appear to be strongly 
dependent on the deposition technology \cite{Terkovsky}.
Amorphous silicon has been indicated as a candidate coating material for third generation (3G) cryogenic detectors 
using crystalline silicon for the mirror  substrate (presently made of fused silica), and a 1550nm laser source \cite{Schnabel} \cite{foot1}

Silicon nitrides, ($\mbox{SiN}_x$) have also been proposed as potentially interesting materials, in view of their flexible stoichiometry,
which allows one to tune their refractive index in a wide range of values; their
ability to accommodate large substrates, via plasma-enhanced chemical vapor deposition (PECVD);
and their fairly low mechanical loss angles, down to $10^{-4}$ and below, at ambient as well as cryogenic temperatures \cite{Pan}, \cite{SiNx_Losses}. 
Their optical losses, though, exceed those of currently used materials \cite{Granata}. 

In Refs. \cite{Yam} and \cite{Steinlechner} it was first suggested to modify the simplest binary coating design 
consisting of alternating  quarter-wavelength (QWL) thick layers  
of $\mbox{SiO}_2$ and 
$\mbox{TiO}_2 ::\mbox{Ta}_2\mbox{O}_5$,
by using a third denser but optically lossier material in the high-index layers closest to the substrate, 
where the field intensity is usually low enough to make its  relatively large optical losses irrelevant.
This would allow one to reduce the total number of layers (and  hence the coating thermal noise), in view of the
larger optical contrast with silica  and/or the lower mechanical losses compared with titania-doped tantala.
The feasibility of $a\mbox{Si}$-based ternary coatings has been recently demonstrated \cite{aSi_prototype}.

Optical coatings using more than two lossy materials ($m$-ary coatings, with $m > 2$) have been studied for a long time (see, e.g. Ref. \cite{Larruquert} for a review).  However, the design constraints and requirements of the mirrors
used by interferometric detectors of gravitational waves are peculiar, especially regarding the key figure
of merit represented by thermal (Brownian) noise, making further analysis necessary.

In this paper we implement exhaustive simulations aimed at identifying the structure of the optimal
ternary coating design yielding the lowest thermal noise under prescribed 
(upper) bounds on power transmittance and absorbance, 
without any {\it a priori} assumption, except that all layers are QWL.

The paper is organized as follows.
In Section \ref{sec:Model} we summarize the relevant modeling assumptions used;
in Section \ref{sec:scrutiny} we introduce the exhaustive procedure used to find the optimum material sequences;
in Section \ref{sec:experiments} we present and discuss the simulations done 
referring to two {\it putative} third materials, 
and in Section \ref{sec:realistic} we present results obtained for realistic materials, namely,
amorphous silicon and silicon nitrides, 
at both ambient and cryogenic temperatures.   
The structure of the coating design families that comply with the prescribed transmittance and absorbance constraints, the properties of the minimum thermal noise (optimal) ones, including robustness against layer thickness errors, and uncertainties or fluctuations in the third material extinction coefficient are illustrated. 
Possible margins for further thermal noise reduction by layer thickness optimization, 
and strategies to implement it, are also discussed.  
Conclusions follow in Section \ref{sec:conclusions}.

\section{COATING MODEL}
\label{sec:Model}

In this section we summarize the modeling assumptions  used throughout this paper,  based on 
the transmission matrix formalism (see, e.g. Refs. \cite{Orfanidis, HarryBook}) 
and the simplest thermal noise model introduced in Ref. \cite{Harry2006}.

\begin{figure}[h!]
\vspace*{-2cm}
\includegraphics[width=9cm]{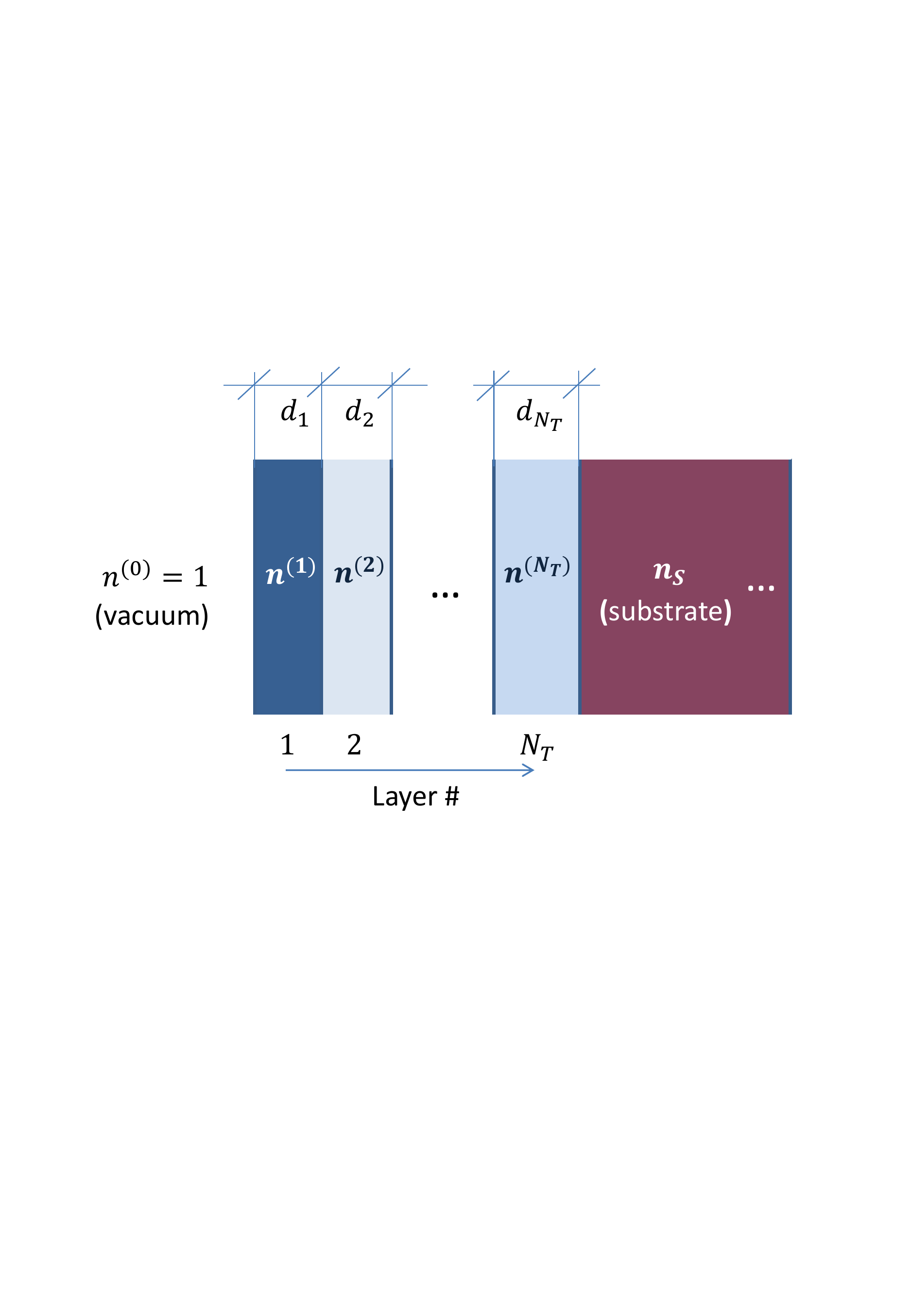}
\vspace*{-5.5cm}
\caption{Sketch of a multilayer coating.}
\label{fig:sketch}
\end{figure}

\subsection{Optical Modeling}

Here, an $\exp(\imath 2 \pi f_0 t)$ time-dependence is understood, $f_0$ being the laser light frequency
and $\imath$ being the imaginary unit.
The optical properties of a multilayer coating can be deduced from its characteristic matrix
\beq
\mathbf{T}=\mathbf{T}_1 \cdot \mathbf{T}_2\cdot \dots \cdot \mathbf{T}_{N_T}
\label{eq:charmat}
\eeq
where $N_T$ is the total number of layers (numbered from the vacuum to the substrate as in Figure \ref{fig:sketch} 
and $\mathbf{T}_m$ is the transmission matrix of the $m$-th layer. Assuming normal incidence \cite{Orfanidis},
\begin{equation}
\mathbf{T}_m=
\left [
\begin{array}{c c}
\cos\left(\psi_m \right)&  (\imath /n^{(m)}) \sin\left(\psi_m\right)\\[10pt]
\displaystyle \imath n^{(m)} \sin\left(\psi_m\right)&
\cos\left(\psi_m\right)
\end{array}
\right ],
\label{eq:Tmatrix}
\end{equation}
where 
\beq
\psi_m=\frac{2 \pi}{\lambda_0}n^{(m)}d_m,
\label{eq:optwidth}
\eeq
$\lambda_0$ and $d_m$ being the light free-space wavelength and the m-th layer thickness, respectively, and
\beq
n^{(m)} = n^{(m)}_r - \imath \kappa^{(m)},
\eeq
$n_r^{(m)}$ being the real refractive index and $\kappa^{(m)}$ being the extinction coefficient of the m-th layer material.

The coating is placed between two homogeneous non dissipative dielectric half-spaces
with refractive indices $n^{(0)}$ and $n_S$, respectively (see Figure \ref{fig:sketch}).
The bottom half-space is the substrate; the top is assumed to be the vacuum, with $n^{(0)}=1$.

The effective complex refractive index of the whole substrate-terminated coating is,
\beq
n_C=\frac{T_{21}+n_S T_{22}}{T_{11}+n_S T_{12}},
\eeq
which can be used to compute the (monochromatic plane wave, normal incidence) coating reflection coefficient $\Gamma_C$
\beq
\Gamma_C=\frac{1-n_C}{1+n_C}
\eeq
and the power transmittance
\beq
\tau_C=\frac{\mathcal{P}_{\mbox{in}}}{\mathcal{P}^{+}}=1 - |\Gamma_c|^2,
\label{eq:transmF}
\eeq
where $\mathcal{P}_{\mbox{in}}$ is the power density flowing into the coating through the vacuum/coating interface and
$\mathcal{P}^{+}$ is the power density of the incident wave,
\beq
\mathcal{P}^+= \frac{1}{2 Z_0}|E_{\mbox{inc}}|^2,
\eeq
where $E_{\mbox{inc}}$ is the (transverse) incident electric field at the vacuum/coating interface and 
${\displaystyle Z_0=\sqrt{ \mu_0/\epsilon_0}}$ is the characteristic impedance of the vacuum.

The average power density dissipated in the coating is the difference between $\mathcal{P}_{\mbox{in}}$
and the power density flowing into the substrate  $\mathcal{P}_{\mbox{out}}$.
This latter can be computed as 
\beq
\mathcal{P}_{\mbox{out}}= \frac{1}{2} \mbox{Re}( E^{(S)}H^{(S)*}),
\eeq
where $\mbox{Re}(\cdot)$ gives the real part of its argument
and $E^{(S)}$ and $H^{(S)}$ 
are the (transverse) electric and magnetic fields at the coating/substrate interface, which
are readily obtained from the fields 
$E^{(0)}=E_{\mbox{inc}} (1+\Gamma_c)$ and
 $Z_0 H^{(0)}= E_{\mbox{inc}} (1-\Gamma_c)$  
 at the vacuum/coating interface  using the formula 
\beq
\left [
\begin{array}{c}
    E^{(S)}\\
   Z_0 H^{(S)}
\end{array}
\right ]
= \mathbf{T}^{-1} 
\left [
\begin{array}{c}
    E^{(0)}\\
    Z_0 H^{(0)}
\end{array}
\right ].
\eeq
Accordingly, the coating absorbance is
\beq
\alpha_C = \frac{(\mathcal{P}_{\mbox{in}}-\mathcal{P}_{\mbox{out}})}{\mathcal{P}^+} =  \tau_C - \tau_S,
\label{eq:assorb}
\eeq
where
\beq
\tau_S = \mathcal{P}_{\mbox{out}}/ \mathcal{P}^+
\eeq
is the fraction of the incident power leaking into the substrate. 
Note that eq. (\ref{eq:assorb}) entails the obvious condition $\tau_C \ge \tau_S$.

\subsection{Thermal Noise Modeling}

The frequency-dependent power spectral density $S_{\mbox{coat}}^{(B)}(f)$ of the coating thermal noise can be written
\beq
S_{\mbox{coat}}^{(B)}(f)\propto \frac{T}{w f}\phi_C,
\label{eq:noise}
\eeq
where $f$ is the frequency, $T$ is the (absolute) temperature, $w$ is the (assumed Gaussian) laser-beam waist,
and $\phi_c$ is the coating loss angle. Neglecting higher-order terms stemming from subtler effects  \cite{Hong}, 
this latter can be written \cite{Harry2006}
\beq
\phi_C= \sum_{m=1}^{N_T} \eta_m d_m,
\label{eq:phic}
\eeq
where 
\beq
\eta_{m}= \frac{1}{\sqrt{\pi} w}  \phi_{m} \left( \frac{Y_{m}}{Y_S} + \frac{Y_S}{Y_{m}} \right)
\label{eq:eta}
\eeq
is the specific loss angle (loss angle per unit thickness)  of the material making the $m$-th  layer,
$\phi_{m}$ and $Y_m$ being its mechanical loss angle and  Young's modulus, respectively, and
$Y_S$ being the Young's modulus of the substrate.

According to eq. (\ref{eq:noise}), 
lowering the temperature $T$  would reduce thermal noise \cite{foot2}.
However, in many coating materials, including those currently in use (silica and titania-doped tantala),
mechanical losses peak \cite{Martin2008} \cite{Martin2014} in the range of the cryogenic temperatures of interest 
for next-generation detectors such as the Einstein Telescope  (ET) \cite{ET} and LIGO-Cosmic
Explorer (CE) \cite{LIGO_CE}, 
pioneered by Kamioka Gravitational Wave Detecto (KAGRA) \cite{KAGRA}.

\section{EXHAUSTIVE SCRUTINY OF QWL TERNARY COATINGS}
\label{sec:scrutiny}

Exhaustive scrutiny  of QWL ternary coatings consists in evaluating the performance (in terms of power transmittance, power absorbance and coating loss angle) of all (admissible) ternary coatings consisting of QWL layers made of {\it three} possible materials,
henceforth denoted as {\it L}, {\it H}, and {\it H'}, that comply with given transmittance and absorbance constraints,
 \beq
 \tau_C \leq \tau_{\mbox{ref}} \mbox{  ,  }\alpha_C \leq  \alpha_{\mbox{ref}}.
 \label{eq:constraints}
 \eeq 
For coatings consisting of QWL layers, the matrices $\mathbf{T}_m$, $m=1,2,\dots,N_T$, 
in (\ref{eq:Tmatrix}) take the simple form \cite{foot3}
%
\beq
\mathbf{T}_m=
\imath
\left[
\begin{array}{lr}
\sinh( \frac{\pi}{2}\frac{\kappa^{(m)}}{n_r^{(m)}})     &  
 \frac{1}{n^{(m)}} \cosh( \frac{\pi}{2}\frac{\kappa^{(m)}}{n_r^{(m)}})
 \\ \\
n^{(m)} \cosh( \frac{\pi}{2}\frac{\kappa^{(m)}}{n_r^{(m)}})    & 
\sinh( \frac{\pi}{2}\frac{\kappa^{(m)}}{n_r^{(m)}})   
\end{array}
\right].
\label{eq:TQWL}
\eeq
For ternary coatings,  $n^{(m)}$ can only take values in  $\left\{n_L, n_H, n_{H'} \right\}$, and the corresponding single-layer matrices will be
denoted as  $\mathbf{T}_{L}$, $\mathbf{T}_{H}$ and $\mathbf{T}_{H'}$, respectively.

Among all possible material sequences, those for which $n^{(m)}=n^{(m-1)}$ for some $m$ should be obviously discarded \cite{foot4}.
%
Accordingly,  we are left with a total of  $N_C=3\times 2^{N_T-1}$  distinct acceptable ternary coatings  
consisting of $N_T$ QWL layers \cite{foot5}.

Knowledge of the matrix (\ref{eq:charmat}) yields the coating transmittance and absorbance, as shown in Sect. \ref{sec:Model}.

In order to compute the coating thermal noise it is expedient to let:
\onecolumngrid
\beq
\gamma_H = 
\frac{\mbox{Re}[n_L]}{\mbox{Re}[n_H]} \frac{\eta_{H}}{\eta_{L}} = 
\frac{\mbox{Re}[n_L]}{\mbox{Re}[n_H]} \frac{\phi_{H}}{\phi_{L}} \left( \frac{Y_{H}}{Y_s} + \frac{Y_s}{Y_{H}} \right)\left( \frac{Y_{L}}{Y_s} + \frac{Y_s}{Y_{L}} \right)^{-1}
\eeq
\beq
\gamma_{H'} = 
\frac{\mbox{Re}[n_L]}{\mbox{Re}[n_{H'}]} \frac{\eta_{H'}}{\eta_{L}}= 
\frac{\mbox{Re}[n_L]}{\mbox{Re}[n_{H'}]} \frac{\phi_{H'}}{\phi_{L}} \left( \frac{Y_{H'}}{Y_s} + \frac{Y_s}{Y_{H'}} \right)\left( \frac{Y_{L}}{Y_s} + \frac{Y_s}{Y_{L}} \right)^{-1}
\eeq
\twocolumngrid
where $\phi_{L}$,  $\phi_{H}$ and $\phi_{H'}$ are the material mechanical loss angles. Hence,
using (\ref{eq:phic}) and (\ref{eq:eta})
\beq
\phi_C= \frac{\eta_L\lambda_0}{4 \mbox{Re}(n_L)} 
\left(N_L + \gamma_H N_H + \gamma_{H'}N_{H'} 
\right)
\eeq
$N_L$, $N_H$ and $N_{H'}$ being the number of layers made of the $L$, $H$, and $H'$ materials, respectively.

The above is a typical constrained optimization problem \cite{CSP}, and has combinatorial complexity.
In order to keep the computational burden and computing times within acceptable limits, 
we use the backtracking strategy \cite{Knuth} to reduce the number of matrix multiplications. 

\section{NUMERICAL EXPERIMENTS}
\label{sec:experiments}
  
In this Section we apply the above mentioned exhaustive scrutinizing procedure to ternary QWL coatings laid
on a fused-silica substrate (assumed to be of infinite thickness), 
using  SiO$_2$ and $\mbox{TiO}_2::\mbox{Ta}_2\mbox{O}_5$  
for the low-index ($L$) and high-index ($H$)  materials, respectively. 
For illustrative purposes, we shall consider first two {\it putative} candidates for the third (high-index) material  ($H'$).
These will be referred to as Material-A and Material-B and represent two rather extreme paradigms,
similar to those discussed in Ref. \cite{Yam}.
 Numerical experiments based on realistic materials, namely, $a\mbox{Si}$ and $\mbox{SiN}_x$, 
 are presented in Section \ref{sec:realistic}.
 
Material-A has the same mechanical losses as $\mbox{TiO}_2::\mbox{Ta}_2\mbox{O}_5$ and
a fairly higher refractive index, but it has larger optical losses;
Material-B has the same refractive index as $\mbox{TiO}_2::\mbox{Ta}_2\mbox{O}_5$ and 
fairly lower mechanical losses,  but it has larger optical losses.
The numerical values of the relevant properties of putative materials A and B 
are collected in Table-I.
We consider different values of the extinction coefficient ranging from $10^{-6}$ to $10^{-4}$.
%
\begin{table}[h!]
\resizebox{\columnwidth}{!}{
\begin{tabular}{|l|l|l|l|l|}
\hline \hline
Property          & SiO$_2$                                 & TiO$_2$::Ta$_2$O$_5$               & MA                        & MB          \\ \hline \hline
$n_r$            & $1.45$                                   & $2.1$                                     & $3.0$                 & $2.1$    \\ \hline
$\kappa$  &     $10^{-11}$                          & $2 \times  10^{-8}$    & $10^{-6}$ \, to \, $10^{-4}$ & $10^{-6}$  \, to \, $10^{-4}$    \\  \hline
$Y$                     & $72$ GPa                                 & $140$ GPa                                 & $100$ GPa        & $100$ GPa    \\ \hline
$\phi$   & $5.0 \times  10^{-5}$ & $3.76 \times  10^{-4}$ & $3.76 \times  10^{-4}$                              & $1 \times  10^{-4}$  \\ \hline \hline                             
\end{tabular}}
\vspace*{0cm}
\caption{Numerical values of relevant parameters for putative materials A and B. 
All symbols have the usual meaning.}
\label{tab:Table_I}
\end{table}

%
%

\onecolumngrid

\begin{table}[p]
\centering
\includegraphics[width=15cm]{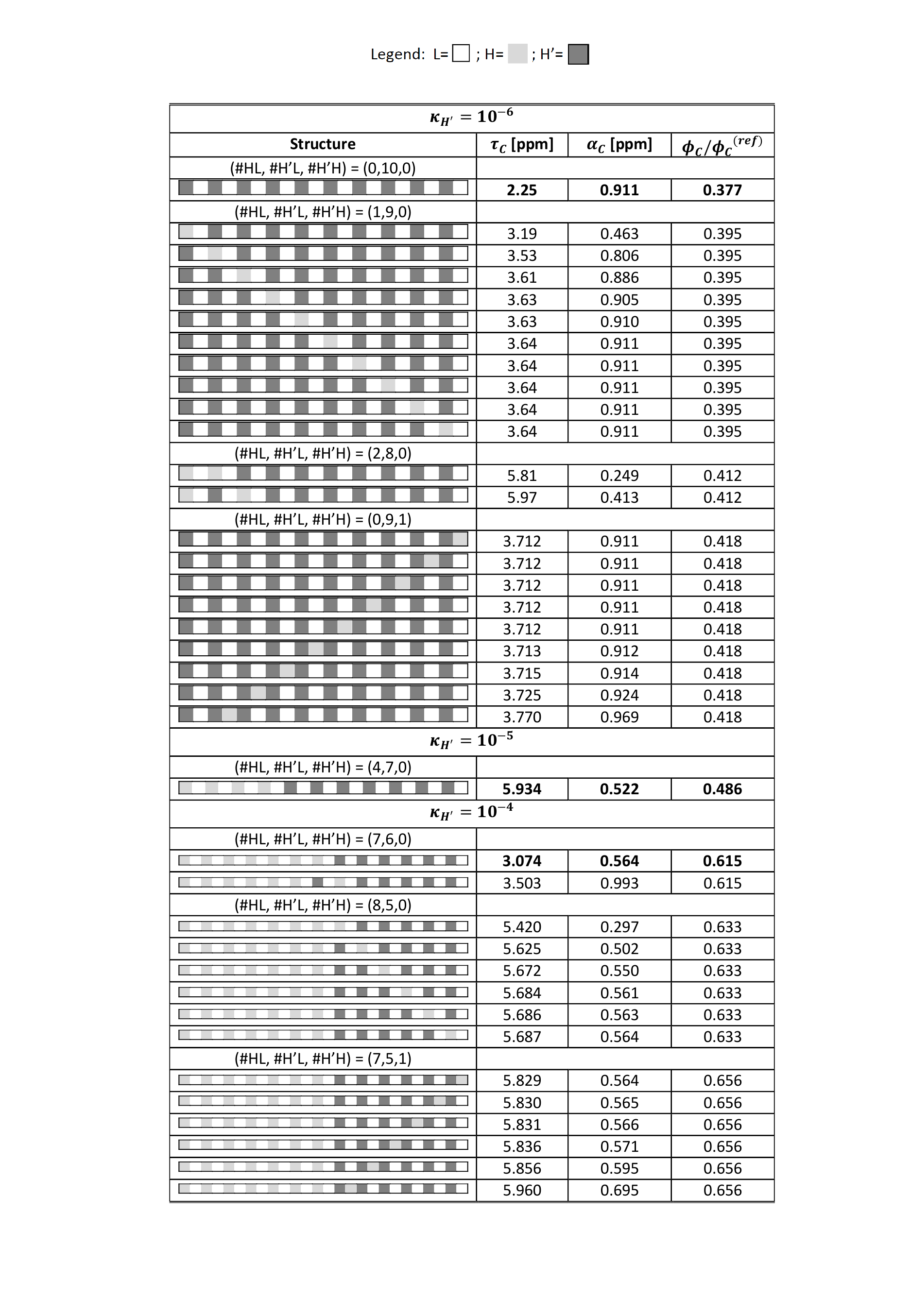}
\caption{The set of all admissible ternary QWL coatings  subject to (\ref{eq:constraints}) using
silica ($L$), titania-doped tantala ($H$) and material A for ($H'$), with $\kappa_{H'}=10^{-6},10^{-5},10^{-4}$. See Section \ref{sec:scrutiny} for details.}
\label{tab:Table-II}
\end{table}

\twocolumngrid
We use the following bounds in the transmittance and absorbance constraints (\ref{eq:constraints}):
\beq
\tau_{ref} = 6 \mbox{ ppm} \mbox{   ,   }
\alpha_{ref}=1 \mbox{ ppm},
\label{eq:bounds}
\eeq
and scale the loss angle of the various admissible solutions to that of a reference LIGO/Virgo-like design,
consisting of $N_T=36$ alternating titania-doped-tantala/silica layers, 
for which
\beq
\phi_C=\phi_C^{(\mbox{ref})}=18  \frac{\eta_L\lambda_0}{4 \mbox{Re}(n_L)} 
\left( 1 + \gamma_H \right).
\eeq
%
                                              
\subsection{Ternary QWL Coatings Using Material-A}

The set of all admissible  ternary QWL coatings compliant with the prescribed transmittance and absorbance 
constraints  (\ref{eq:bounds}) and using Material-A for $H'$ is collected in Table-II  
for three possible values of the extinction coeffcient ($\kappa_{H'}=10^{-6},10^{-5},10^{-4}$).

They are conveniently grouped  into subsets  featuring the same number of $[H|L]$, $[H'|L]$, and $[H'|H]$ doublets,
and hence the same coating loss angle, in order of increasing  transmittance and/or absorbance.
For all considered $\kappa_{H'}$ values, the optimal design featuring the lowest coating loss angle
consists of  a stack of  $N_{H'}$  doublets $[H'|L]$ grown on top of the substrate, 
topped by another stack of $N_H$  doublets $[H|L]$, 
where the optimal values of $N_{H'}$ and $N_H$ depend on the  extinction coefficient of the $H'$ material.

There are no $[H'|H]$  (or $[H|H']$)  doublets in the optimal designs.

These findings confirm the  heuristic assumption first made in Refs. \cite{Yam} and \cite{Steinlechner} about the
structure of ternary QWL coatings yielding minimal noise under prescribed transmittance and absorbance
constraints.

\subsection{Ternary QWL Coatings Using Material-B}

In the case of ternary QWL coatings using Material-B,, materials $H$ and $H'$ are iso-refractive; hence the total number of high-low index doublets
needed to satisfy the prescribed transmittance constraint remains fixed, irrespective of whether the 
high-index
layers consist of the $H$ or $H'$ material, 
and is the same as for the  reference TiO$_2$::Ta$_2$O$_5$/SiO$_2$  binary coating. Hence, 
\beq
N_H + N_{H'} = N_L=N_L^{(\mbox{ref})}.
\eeq
Note also that in this case, $[H'|H]$  (or $[H|H']$)  doublets are forbidden, being optically homogeneous
and half-wavelength thick.

\begin{figure}[h!]
\hspace*{-2cm}
\includegraphics[width=12cm]{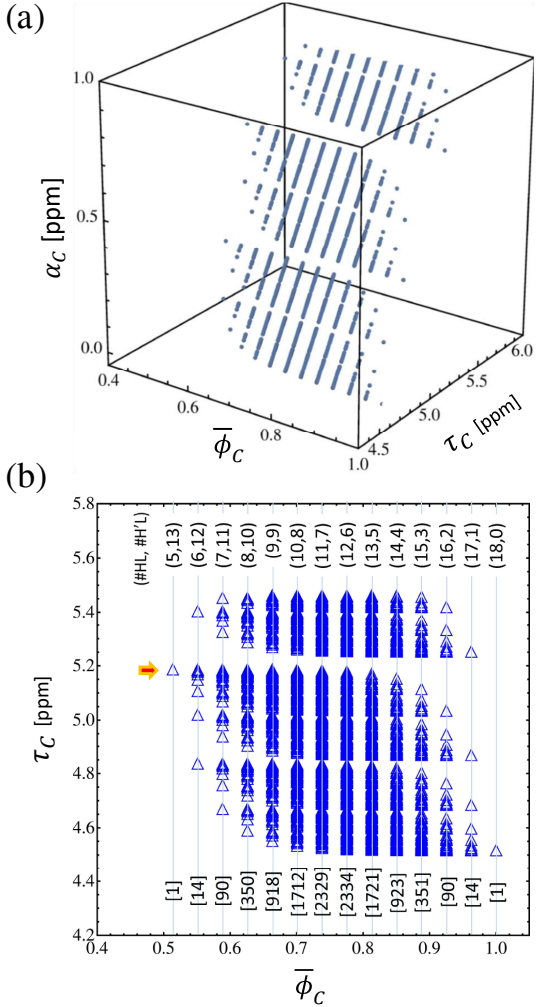}
\caption{The set of all admissible ternary QWL coatings  subject to (\ref{eq:constraints}) using
silica ($L$), titania-doped tantala ($H$) and Material-B for ($H'$), with $\kappa_{H'}=10^{-5}$
and $\bar{\phi}_C=\phi_C/\phi^{(\mbox{ref})}_C$. See Section \ref{sec:scrutiny} for details.}
\label{fig:All_B}
\end{figure}

The set of all admissible ternary QWL coatings compliant with the prescribed transmittance and absorbance 
constraints  (\ref{eq:bounds}) can be conveniently visualized  as in Figure \ref{fig:All_B},
which refers to the case  $\kappa_{H'}=10^{-5}$.
%
%

It is seen that all sub-optimal, constraint-compliant admissible designs can be divided  into distinct families, represented
by the aligned markers in Figure \ref{fig:All_B} (a) and \ref{fig:All_B}(b), 
each family  featuring a number ($N_H$) of  $[H|L]$  doublets 
, denoted as $\#HL$ in figure \ref{fig:All_B} (b), 
and a fixed number ($N_{H'}=N_L-N_H$) of $[H|'L]$ doublets,
denoted as $\#H'L$ in figure \ref{fig:All_B} (b);
hence each family features the same loss-angle but different transmittances and absorbances,
as seen from Figure \ref{fig:All_B} (a).

For each family,  the number of distinct admissible designs, in square brackets in Figure \ref{fig:All_B} (b) 
is (slightly) less than the binomial coefficient
\beq 
\left(   
\begin{array}{c}
N_H
\\
N_L-N_H
\end{array}
\right)
\eeq
due to the (relatively few) designs that do not fulfill the transmittance and absorbance constraints.

Similar to the previous case, the optimal design featuring the lowest thermal noise
under the prescribed transmittance and absorbance constraints, 
consists of  a stack of $[H'|L]$  doublets  grown on top of the substrate, 
topped by another stack of $[H|L]$  doublets, again confirming the ansatz in  Refs. \cite{Yam} and \cite{Steinlechner} .

\begin{figure}[h!]
\includegraphics[width=9cm]{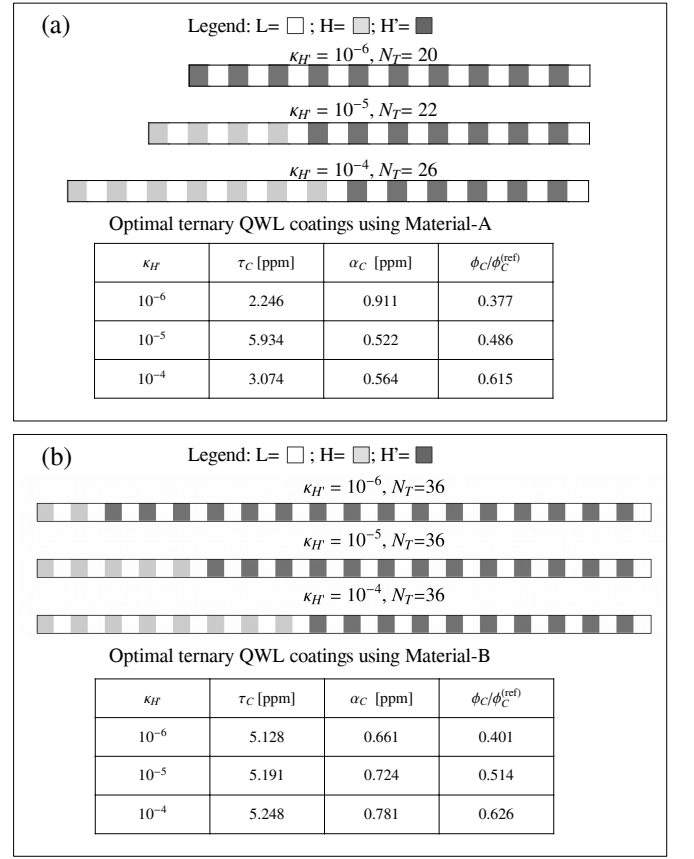}
\caption{The found optimal (minimum Brownian noise) QWL ternary coating designs subject to (\ref{eq:constraints}), 
using silica ($L$), titania-doped tantala ($H$) and materials A and B ($H'$), assuming different values for $\kappa_{H'}$
(substrate on the right).}
\label{fig:optimal}
\end{figure}

The found optimal designs using materials A and B for $H'$, with $\kappa_{H'}=10^{-6},10^{-5},10^{-4}$,
are displayed  in Figure \ref{fig:optimal} (a) and \ref{fig:optimal} (b), respectively.
%

\subsection{Robustness}

The optimal ternary QWL coatings are nicely robust against uncertainties in the value of the extinction coefficient $\kappa_{H'}$, as well as 
against unavoidable inaccuracies in the layers' thicknesses, due to technological limitations of the deposition process \cite{foot6}.

\begin{figure}[h!]
\hspace*{-2cm}
\includegraphics[width=11cm]{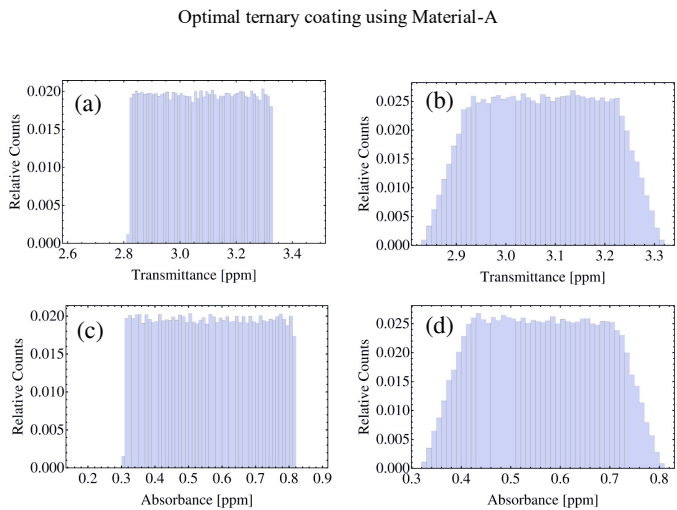}
\vspace*{-6cm}
\caption{Distributions of coating transmittance and absorbance in a sample of  $10^5$ realizations  
of the optimal ternary QWL coating using Material-A,  assuming $\kappa_A$ to be random  uniform in
$(0.5\bar{\kappa}_A , 1.5\bar{\kappa}_A)$ , with $\bar{\kappa}_A=10^{-4}$.
Panels (a), (c) refer to the case where the extinction coefficient is the same for all  $H'$ layers; panels (b),
(d) to the case ii) where the extinction coefficients of the $H'$ layers are {\it independent} identically distributed
random variables.
}
\label{fig:A_randomK}
\end{figure}

As an illustration, Figure \ref{fig:A_randomK}  shows the distributions of coating transmittances and absorbances 
in  $10^5$ realizations of the optimal ternary QWL coating using Material-A.
%

Figures \ref{fig:A_randomK}(a) and \ref{fig:A_randomK}(c) refer to  case (i), where  the extinction coefficient $\kappa_A$  is the same for all $H'$
layers and is random uniformly distributed in in $(0.5\bar{\kappa}_A , 1.5\bar{\kappa}_A)$, with $\bar{\kappa}_A=10^{-4}$; 

Figs. \ref{fig:A_randomK}(b) and \ref{fig:A_randomK}(d) refer to case (ii), where the extinction coefficients of the $H'$ layers 
are {\it independent} random variables, identically distributed as in case (i).

\begin{figure}[h!]
\hspace*{-1.5cm}
\includegraphics[width=11cm]{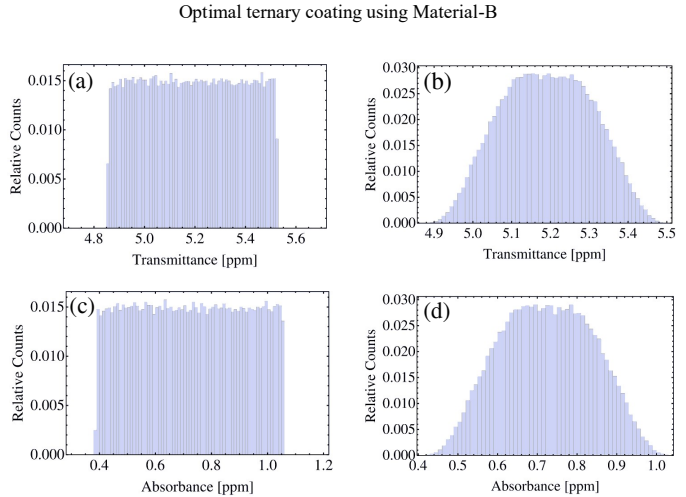}
\vspace*{-6cm}
\caption{Distributions of coating transmittance and absorbance in a sample of  $10^5$ realizations  
of the optimal ternary QWL coating using Material-B,  assuming $\kappa_B$ to be random  uniform in
$(0.5\bar{\kappa}_B , 1.5\bar{\kappa}_B)$ , with $\bar{\kappa}_B=10^{-5}$.
Panels (a), (c) refer to the case where the extinction coefficient is the same for all $H'$ layers; panels
(b), (d) refer to the case ii) where the extinction coefficients of the $H'$ layers are {\it independent} identically
distributed random variables.
}
\label{fig:B_randomK}
\end{figure}

Similarly, Figure \ref{fig:B_randomK}  shows the distributions of coating transmittance and absorbance 
in a sample of  $10^5$ realizations  of the optimal ternary QWL coating using Material-B, 
assuming $\kappa_B$ to be random uniformly distributed in  $(0.5\bar{\kappa}_B , 1.5\bar{\kappa}_B)$, with $\bar{\kappa}_B=10^{-5}$.

Also in this case, Figs. \ref{fig:B_randomK}(a) and \ref{fig:A_randomK}(c) refer to the case where the random extinction coefficient is the same for all $H'$ layers; Figs.
5(b) and 5(d) refer to the case where the extinction coefficients of the $H'$ layers are independent and
identically distributed as specified.
%
%

\begin{figure}[h!]
\hspace*{-1cm}
\includegraphics[width=11cm]{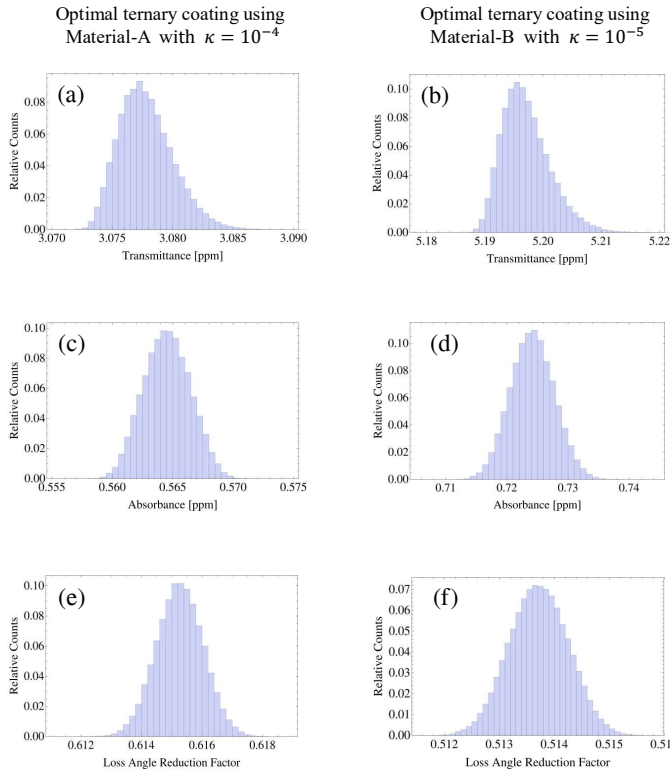}
\vspace*{-2cm}
\caption{Distributions of coating transmittance, absorbance and loss angle 
(normalized to the value of the reference binary coating) in a sample of $10^5$  realizations,
assuming the thicknesses of all layers to be independent  random variables identically distributed 
uniformly around the nominal QWL thickness, in a symmetric interval of total width $2$ nm.
(a), (c), and (e) Optimal ternary QWL coating using Material-A for $H'$,  with $\kappa_A=10^{-4}$.
(b), (d), and (f) Optimal ternary QWL coating using MaterialB for $H'$,  with $\kappa_B=10^{-5}$.}
\label{fig:random_d}
\end{figure}

Not unexpectedly, uncertainties in the extinction coefficient stemming from fluctuations in the deposition process,
rather than systematic uncertainty in the nominal value, have a lesser effect, due to possible fluctuation compensation,
resulting in narrower distributions of the coating transmittance and absorbance.    
\\
Figure \ref{fig:random_d} shows the distributions of coating transmittance, absorbance, and loss angle 
(normalized to the value of the reference binary coating) in a sample of $10^5$  realizations of
(i) the optimal ternary QWL coating using Material-A for $H'$,  
with $\kappa_A=10^{-4}$ [Fig. 6(a), 6(c) and 6(e)], and
(ii) the optimal ternary QWL coating using Material-B for $H'$,  with $\kappa_B=10^{-5}$ [Figs. 6(b), 6(d), and 6(f)],
assuming the thicknesses of all layers to be independent  random variables identically distributed 
uniformly around the nominal QWL thickness, in a symmetric interval of total width $2\,\,\mbox{nm}$.  
%
%

We may conclude that the optimal ternary QWL designs are fairly robust against uncertainties in the
extinction coefficient of the $H'$ material, and deposition-related thickness errors. 

\subsection{Transmittance Spectra}

\begin{figure}[h!]
\hspace*{-1cm}
\includegraphics[width=11cm]{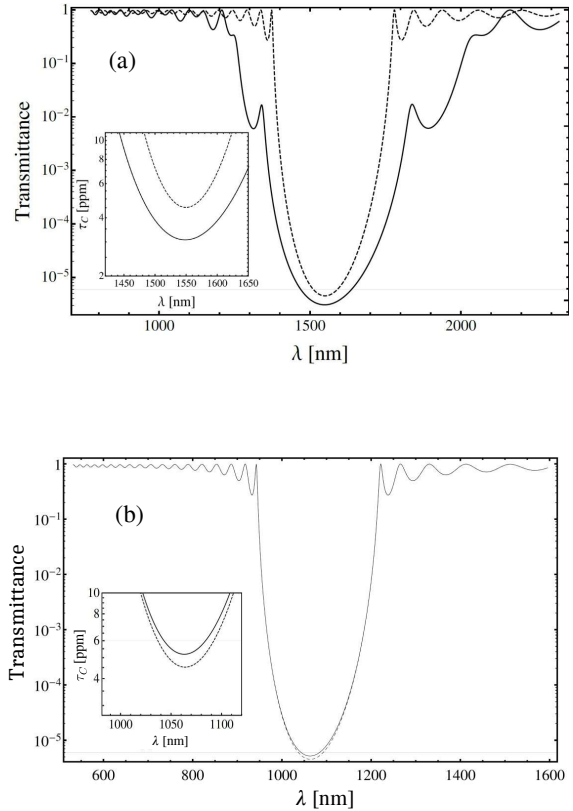}
\vspace*{-2cm}
\caption{
Transmittance spectra of optimal designs using  Material-A with $\kappa_{H'}=10^{-4}$, and Material-B
with $\kappa_{H'}=10^{-5}$  respectively in (a) and (b), for  material $H'$. 
Close-ups are shown in insets.
}
\label{fig:tau_spectra}
\end{figure}

We computed the  transmittance spectra of the above optimal  ternary QWL coatings.
These are shown in Figs. \ref{fig:tau_spectra} (a) and \ref{fig:tau_spectra} (b) for coatings using materials A and B, respectively.
%
%

The spectra were computed neglecting chromatic dispersion, except in the neighborhood of the operating
wavelength ($\lambda_0= 1064$ nm), shown in the insets, where a linear approximation was used,
 
\beq
n_r(\lambda)=n_r(\lambda_0) + \left. \frac{dn_r}{d\lambda}\right|_{\lambda_0} (\lambda - \lambda_0).
\label{eq:chroma}
\eeq
The pertinent values of  $\left. \frac{dn_r}{d\lambda}\right|_{\lambda_0}$ were taken from 
Ref. \cite{slopes} and are
$-1.2 \cdot 10^{-5}\mbox{ nm}^{-1}$ for silica,  
$-4.9 \cdot 10^{-5}\mbox{ nm}^{-1}$ for titania-doped tantala, 
$-3.8 \cdot 10^{-5} \mbox{ nm}^{-1}$ for Material-A,
and $-4.26 \cdot 10^{-4} \mbox{ nm}^{-1}$ for Material-B. 

Within the limits of this model, no ripple is observed in the high-reflectance band.
The shape of the transmission spectrum for the optimal coating using Material-A departs more markedly
from the reference spectrum  compared with that of the optimal coating using Material-B.
The observed asymmetry of the lobes stems from the fact that the coating is piecewise homogeneous,
consisting of two cascaded homogeneous QWL stacks.

\subsection{Thickness Optimization}

Thermal noise in binary coatings can be effectively reduced  
compared with the  reference QWL-layer design 
by suitably reducing the total thickness of the mechanically noisier material(s), while increasing the 
total thickness of the other  material(s) and the total number of doublets, so as
to keep the coating transmittance unchanged \cite{HarryBook}.  

Remarkably,  the thickness-optimized binary coatings turn out to consist of almost identical 
stacked $[H|L]$ doublets whose thickness is one half of the working wavelength (Bragg condition),
the exception being represented by a few layers near the coating top and bottom
\cite{Agresti}, \cite{Villar}, \cite{Gurkovsky}, \cite{Pierro}.

Implementing thickness optimization for ternary coatings is computationally demanding.
In a full-blind exhaustive approach, each and any $L$ layer should be allowed to take 
any thickness value in the range ($\lambda/4,\lambda/2$),
and each and any $H$ and $H'$  layer should be allowed to take
any thickness in ($0,\lambda/4$), $\lambda$ being the local wavelength.
Even after suitable discretization of the above search intervals, the computational burden of 
an {\it exhaustive} search would become prohibitive for any meaningful value of $N_T$.

A reasonable heuristic approach to thickness optimization of ternary QWL coatings may thus consist in optimizing 
 the two binary QWL stacks  that form the top and bottom parts of the optimal QWL ternary coatings,
assuming each of them to consist of identical non-QWL Bragg doublets \cite{Pinto_Triplets}. 

Here, to illustrate the possible margins of further loss angle reduction obtainable from thickness optimization,
we content ourselves with computing the (normalized) coating loss angle, absorbance and power transmittance 
of the coatings obtained after modifying  the optimal ternary QWL coatings  found in the previous Sections
and listed in Figure \ref{fig:optimal} by letting 
\beq
\frac{n_H d_H}{\lambda_0}=\frac{n_H' d_H'}{\lambda_0}=\frac{1}{6} \mbox{    and   } 
\frac{n_L d_L}{\lambda_0}=\frac{1}{3}
\eeq
so as to preserve the Bragg character of the doublets, 
and adding a few $[H|L]$ and/or $[H'L]$ layers as needed to maintain compliance with the transmittance and absorbance constraints.

The transmittance, absorbance, and coating loss angle (normalized to the reference value) of 
the resulting thickness-tweaked coatings are collected in Table-III.

\onecolumngrid

\begin{table}[h!]
\centering
\hspace*{-0.2cm}
\includegraphics[width=18cm,keepaspectratio]{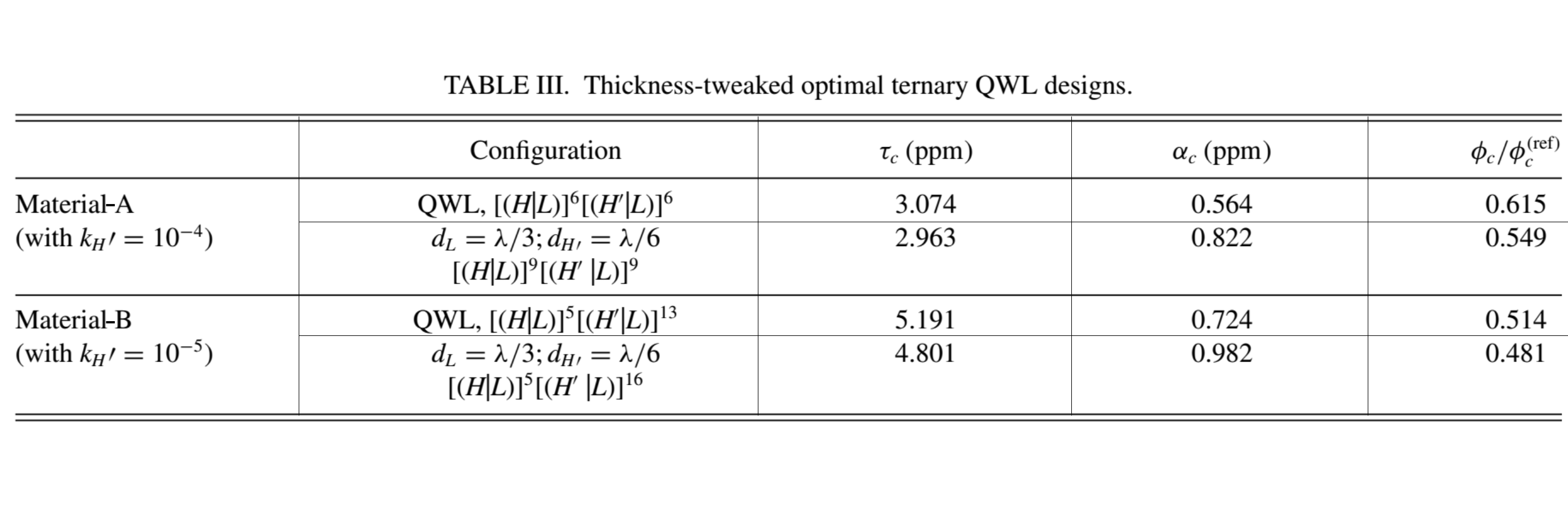}
\vspace*{-0cm}
\label{fig:Table-III}
\end{table}

\twocolumngrid

\vspace*{0.5cm}
\section{Realistic Materials}
\label{sec:realistic}
\begin{figure}[h!]
\includegraphics[width=9cm]{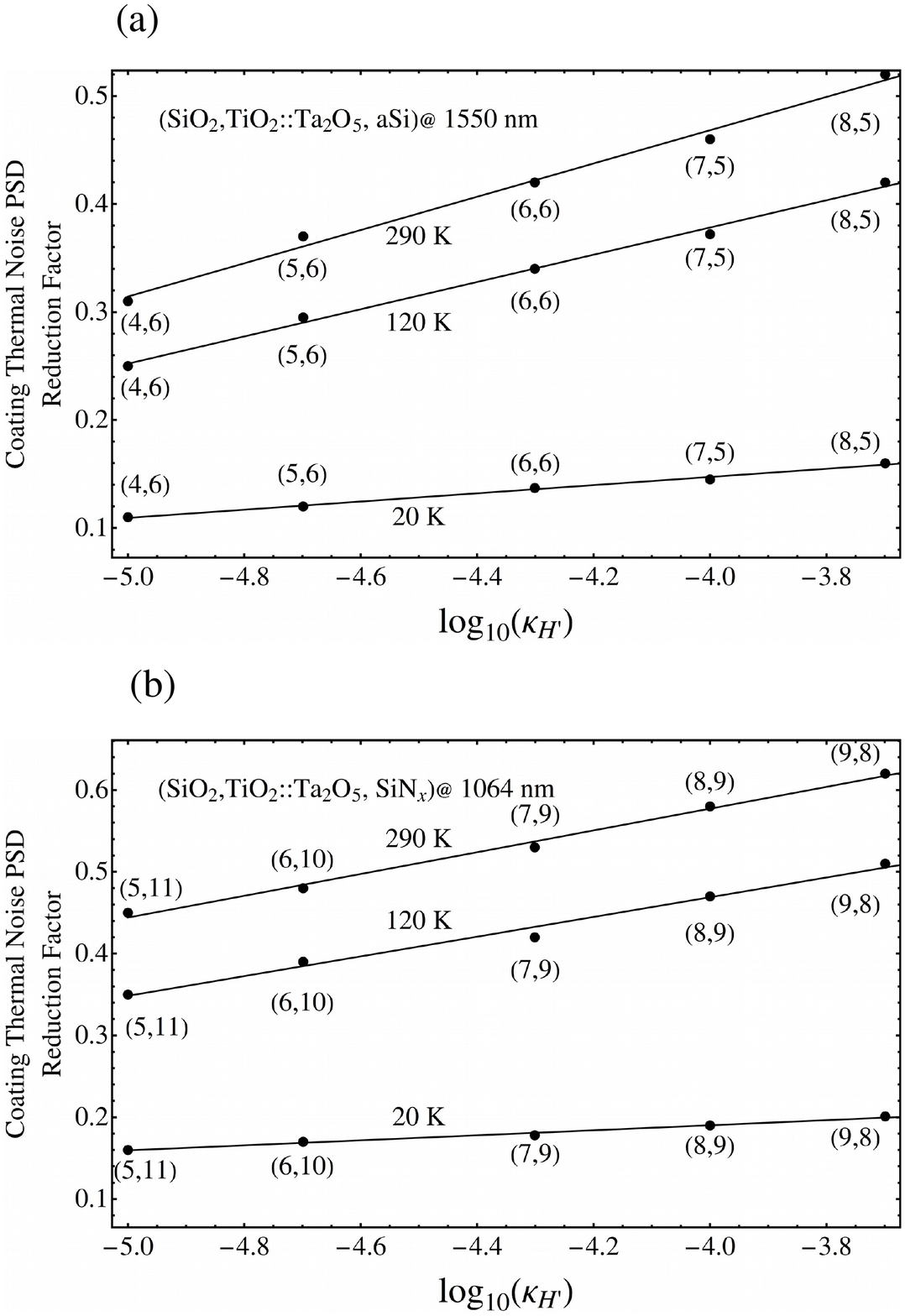}
\caption{Achievable thermal (Brownian) noise reduction factors for optimal QWL ternary coatings 
vs extinction coefficient of third $(H')$ material.
The coatings in (a) use $a\mbox{Si}$  (at 1550nm) and those in (b) use $\mbox{SiN}_x$  (at 1064nm) 
as the third material, together with 
$\mbox{SiO}_2$  and $\mbox{Ta}_2\mbox{O}_5$. Three different operating temperatures (290K, 120K, and 20K) are considered.}
\label{fig:realistic}
\end{figure}

The structure and properties of the optimal coatings are the same if we consider {\it realistic}
candidates for the third material, namely $a\mbox{Si}$, and $\mbox{SiN}_x$.
In this section we present results based on the previous analysis/simulation tools
for optimal QWL ternary coatings operating at 290K, 120K, and 20K,
based i)  on (silica, titania-doped tantala and $a$Si) at 1550nm
and  ii)  on (silica, titania-doped tantala and SiN$_x$) at 1064nm, 
using current available measurements/estimates of the actual 
(or fiducially achievable) material parameters, collected in Table IV.
The coating Brownian noise power spectral density (PSD) reduction factor with respect to the reference advanced LIGO (aLIGO) and advanced Virgo (adVirgo)
coatings currently in operation (at ambient temperature) is shown in Figure 
\ref{fig:realistic}(a) and \ref{fig:realistic}(b),
respectively for different values of the extinction coefficient  of the third material, 
in the range from $10^{-5}$ to $10^{-4}$.
 %
%

\onecolumngrid

\begin{table}[h!]
\hspace*{-0cm}
\includegraphics[width=18cm]{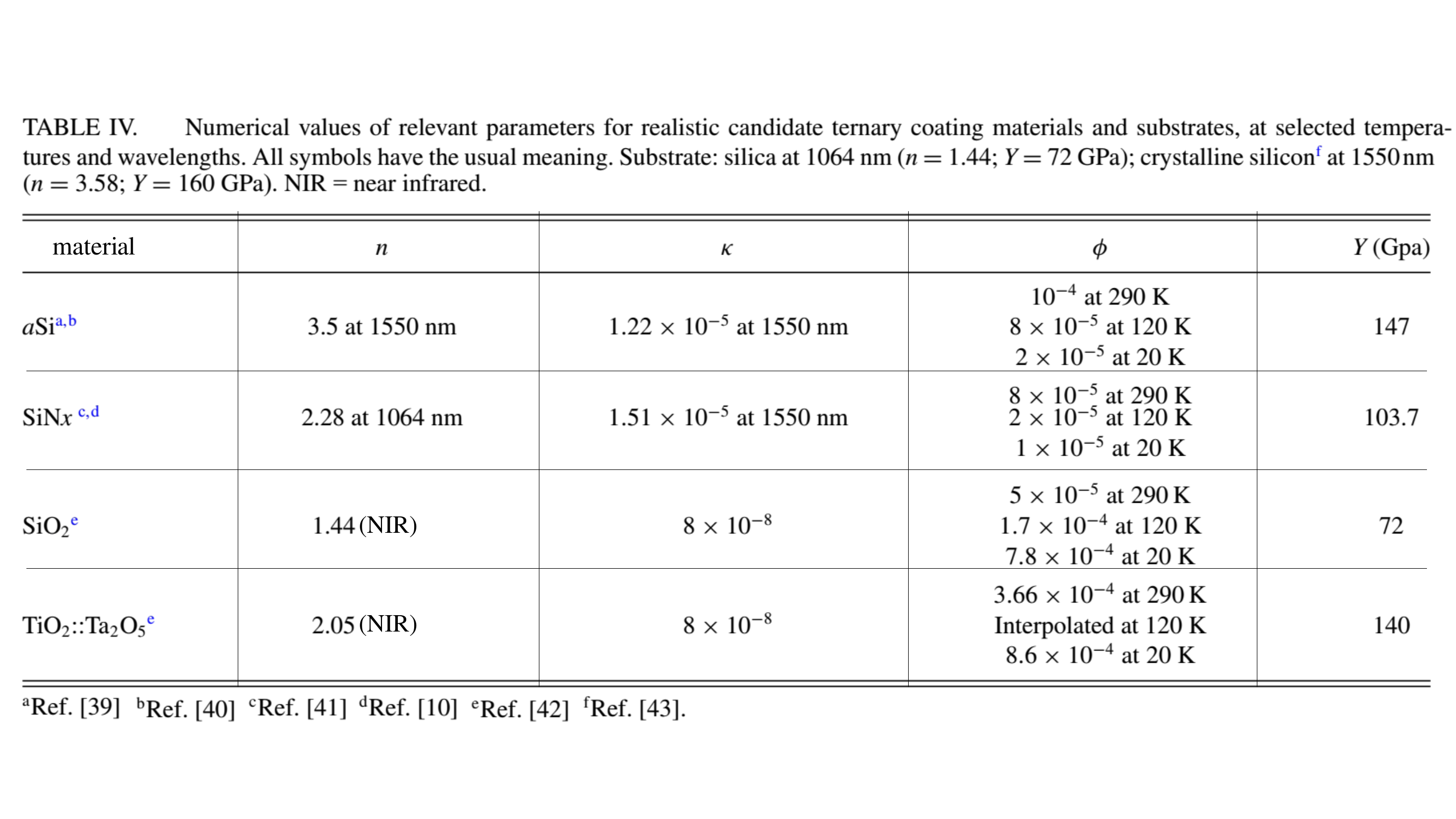}
\vspace*{-1.5cm}
\label{tab:Table-IV}
\end{table}

\twocolumngrid

In calculating the Brownian noise PSD reduction factor we obviously include the temperature-dependent factor in eq. (\ref{eq:noise}).
We note in passing that reducing the PSD by a factor  $\rho$  corresponds to reducing the so called amplitude
spectral density (or rms noise level) by a factor $\rho^{1/2}$ and to boosting the 
visibility distance by a factor 
$\rho^{-1/2}$.
%
%

The quantities $(\#HL,\#H'L)$ in round brackets in Figure \ref{fig:realistic} represent the numbers of $[H|L]$ 
and $[H'|L]$ doublets  in the top and bottom stacks of the found optimal ternary QWL design featuring
the minimal thermal noise under the prescribed constraints 
$\tau_C \leq 6\mbox{ppm}$ and $\alpha_C \leq 1 \mbox{ppm}$.

Remarkably, as seen from the figure, the thermal noise reduction factor is found to be almost linear in $\log_{10}(\kappa_{H'})$.

The achievable coating thermal noise PSD reduction factor is nicely large, especially at cryogenic
temperatures, where it gets close to the Einstein Telescope \cite{ETc} and Cosmic Explorer \cite{CEc} requirements, 
and compares to the expected performance of crystalline coatings \cite{Cole},
while possibly posing less demanding technological challenges.

\section{CONCLUSIONS}
\label{sec:conclusions}

In this paper we addressed the problem of designing a ternary optical coating consisting of QWL layers
to achieve the minimum thermal (Brownian) noise under prescribed optical transmittance and absorbance constraints.

We first considered ternary coatings where two materials are those 
presently in use in the advanced LIGO and Virgo detectors, namely, 
$\mbox{SiO}_2$ and $\mbox{TiO}_2::\mbox{Ta}_2\mbox{O}_5$, 
featuring the best trade-off between optical contrast, optical losses and thermal noise so far;
and the third material is one of two {\it putative}  materials featuring, respectively, 
the same mechanical losses as $\mbox{TiO}_2::\mbox{Ta}_2\mbox{O}_5$ and
a higher refractive index, but larger optical losses (Material-A),
or the same refractive index as $\mbox{TiO}_2::\mbox{Ta}_2\mbox{O}_5$ and 
fairly lower mechanical losses,  but larger optical losses (Material-B),
allowing their extinction coefficient to range from $10^{-6}$ to $10^{-4}$
in both cases.

We performed an exhaustive search over all possible (and admissible,
in the sense discussed in Sect. \ref{sec:experiments}) configurations consisting of QWL layers, using backtracking for numerical efficiency and seeking  the optimal designs
yielding minimum thermal (Brownian) noise under prescribed
upper bounds for transmittance and absorbance.

The main results of this study can be summarized as follows.

All found optimal designs,  consist of  a stack of $[H'|L]$  doublets  grown on top of the substrate, 
topped by another stack of $[H|L]$  doublets, confirming the ansatz in Refs. \cite{Yam,Steinlechner} .

They  are nicely robust against deposition inaccuracies in the individual layer thicknesses and
systematic uncertainties and/or fluctuations from layer to layer of the extinction coefficient.
Their transmittance spectra satisfy the design constraints in the useful band.

We have further shown that a further improvement in performance can be achieved by using thinner layers of
the noisier materials in each Bragg doublet.
Exhaustive blind thickness optimization of ternary coatings appears to be computationally unaffordable, though,
and the problem will be accordingly studied in a future paper  following a heuristic approach.

Next we applied the same optimization strategy to realistic candidates for the third material, 
namely $a\mbox{Si}$ and $\mbox{SiN} _x$, operating at three different temperaures  (290K, 120K, and 20K).
In order to take into account the present uncertainties (and possible margins of improvement \cite{Birney}) 
for the optical losses of these  materials, we considered different values of their extinction coefficient 
in the range from $10^{-6}$ to $10^{-4}$.

All optimal designs outperform significantly the reference binary solution consisting of alternating QWL layers  
made of silica  and titania-doped tantala, in terms of Brownian thermal noise. 

In particular, according to our simulations,  QWL ternary coatings using either $a$Si  or SiN$_x$  in addition to 
silica and titania-doped tantala may achieve an almost tenfold reduction in the coating thermal noise 
(power spectrum)  level compared with  second generation detector coatings operating at ambient temperature.

\section*{ACKNOWLEDGEMENTS}

This work has been supported in part by Istituto Nazionale di Fisica Nucleare (INFN) through the projects Virgo and ET-Italy, and by the European Gravitational Observatory (EGO). 
The authors gratefully acknowledge useful discussions with, and suggestions  from members of the Virgo Coating R \& D Group (VCR \& D) and the LIGO Optics Working Group (OWG).

\clearpage


\begin{thebibliography}{99}
\bibitem{foot1}{The use of $2-\mu$m light sources has also been suggested [47], subject to substantial progress  in the candidate sources (Tm:YAG or Ho:YAG lasers) in terms of 
power-noise trade-off.}
\bibitem{foot2}{
Other options that also appear promising, though on different timescales, including wide beams, crystalline materials, 
nanolayered composites, coating-less mirrors and diffractive optics \cite{Principe} fall beyond the scope
of this paper.}
\bibitem{foot3}{Note that expansion of the hyperbolic functions to first order in  $\kappa^{(m)}$ 
(assumed to be $\ll \, 1$ for all materials involved) may lead to inaccuracies in the computed absorbances that could be non-negligible comparable with
the enforced absorbance bounds.}
\bibitem{foot4}{
Two consecutive QWL layers with the same refractive index would make a half-wavelength optically homogeneous layer, whose transmission matrix (\ref{eq:Tmatrix}) is the identity matrix. Accordingly, such a doublet would not affect coating transmittance, while increasing coating thermal noise.}
\bibitem{foot5}{
This follows immediately by noting that the first layer can be chosen in three possible ways, while each of the $N_T-1$ subsequent ones can be chosen only in two possible ways, 
in view of the condition $n^{(m)} \neq n^{(m-1)}$.}
\end{thebibliography}
\end{document}